\documentclass[sigconf]{acmart}
\usepackage{subfigure}
\usepackage{verbatim}

\AtBeginDocument{%
  \providecommand\BibTeX{{%
    \normalfont B\kern-0.5em{\scshape i\kern-0.25em b}\kern-0.8em\TeX}}}

\setcopyright{acmcopyright}
\copyrightyear{2018}
\acmYear{2018}
\acmDOI{10.1145/1122445.1122456}

\acmConference[FDG '20]{The Fifteenth International Conference on the Foundations of Digital Games}{September 15--18, 2020}{Bugibba, Malta}
\acmBooktitle{The Fifteenth International Conference on the Foundations of Digital Games (FDG '20), September 15--18, 2020, Bugibba, Malta}
\acmPrice{15.00}
\acmISBN{978-1-4503-XXXX-X/18/06}

\newcommand{\hlinesep}{\unskip\ \vrule\ }
\begin{document}

\title{"And then they died": Using Action Sequences for Data Driven, Context Aware Gameplay Analysis}



\author{Erica Kleinman} 
\affiliation{%
  \institution{Northeastern University}
  \city{Boston}
  \state{Massachusetts}
}
\email{kleinman.e@husky.neu.edu}

\author{Sabbir Ahmad} 
\affiliation{%
  \institution{Northeastern University}
  \city{Boston}
  \state{Massachusetts}
}
\email{ahmad.sab@husky.neu.edu}

\author{Zhaoqing Teng} 
\affiliation{%
  \institution{Northeastern University}
  \city{Boston}
  \state{Massachusetts}
}
\email{teng.z@husky.neu.edu}

\author{Andy Bryant} 
\affiliation{%
  \institution{Harvard University}
  \city{Cambridge}
  \state{Massachusetts}
}
\email{A@andymbryant.com}

\author{Truong-Huy D. Nguyen} 
\affiliation{%
  \institution{Google}
  \city{New York}
  \state{New York}
}
\email{truonghuy@gmail.com}

\author{Casper Harteveld} 
\affiliation{%
  \institution{Northeastern University}
  \city{Boston}
  \state{Massachusetts}
}
\email{c.harteveld@northeastern.edu}

\author{Magy Seif El-Nasr} 
\affiliation{%
  \institution{Northeastern University}
  \city{Boston}
  \state{Massachusetts}
}
\email{magy@northeastern.edu}

\renewcommand{\shortauthors}{Kleinman, et al.}

\begin{abstract}
Many successful games rely heavily on data analytics to understand players and inform design. Popular methodologies focus on machine learning and statistical analysis of aggregated data. While effective in extracting information regarding player action, much of the context regarding when and how those actions occurred is lost. Qualitative methods allow researchers to examine context and derive meaningful explanations about the goals and motivations behind player behavior, but are difficult to scale. In this paper, we build on previous work by combining two existing methodologies: \emph{Interactive Behavior Analytics} (IBA) \cite{labelPaper} and sequence analysis (SA), in order to create a novel, mixed methods, human-in-the-loop data analysis methodology that uses behavioral labels and visualizations to allow analysts to examine player behavior in a way that is context sensitive, scalable, and generalizable. We present the methodology along with a case study demonstrating how it can be used to analyze behavioral patterns of teamwork in the popular multiplayer game \emph{Defense of the Ancients 2} (DotA 2).
\end{abstract}



\maketitle


\section{Introduction}

Games have grown exponentially in popularity and accessibility in recent years. This is especially true of esports, an industry worth close to a billion dollars \cite{value}. These online games are frequently updated and adjusted to keep them engaging and relevant. But with so much revenue on the line, it is critical that the developers making these design adjustments are able to do so in an informed manner. In order to make these informed decisions, many have turned to data driven techniques in order to gain better understandings of players' strategic decisions and behavioral patterns \cite{hooshyar2018data}.

Many methods of gameplay analysis exist, with quantitative approaches to data mining and visualization being some of the most popular \cite{wallner2013visualization,hooshyar2018data}. These techniques mostly focus on using algorithms, statistics, and clustering to identify player types, gameplay roles, and indicators of victory \cite{ong2015player,makarov2017predicting,nascimento2017profiling}. While these methodologies are effective at analyzing data efficiently on large scales, they rarely preserve the context from which the data was drawn. This makes it difficult to examine player strategies or problem solving techniques that are very much situated actions \cite{suchman1987plans}, dependent on and informed by the context in which they occur. 

By context we refer to the factors inside the game that directly or indirectly relate to the choices players can make in games. This can include time and location or more detailed factors, such as which characters, items, and environmental features are involved in defining a behavior. Existing techniques, such as heat-maps, are able to account for the more general contextual factors, especially location. However, if one wishes to understand behavioral data in the context that the players consider it, then more detailed information must be captured. For example, existing techniques can show that a kill occurred in a given location at a given time. But if one asked the players involved what happened, they would likely describe an in-game strategic behavior, such as a ``gank", that involved a number of complex details including positioning, environmental objects (rivers and trees), heroes playing specific roles, and a detailed sequence of more granular behaviors that the ``gank" encompassed. Such detailed contextual factors are not well captured by most existing techniques.


If one wishes to examine gameplay in a context aware manner, qualitative methodologies offer this opportunity. Qualitative techniques that have been used to analyze online, multiplayer gameplay include ethnography, interviews, and observation \cite{steinkuehler2004learning,bardzell2008blissfully,kow2013media}. These approaches often involve communicating directly with the players, allowing for context aware analysis. These studies have produced a corpus of information regarding how players self organize, communicate, and collaborate in online gaming environments such that they are successfully able to pursue goals and achieve success \cite{kow2013media,steinkuehler2004learning,bardzell2008blissfully,Kou2014}. However, these techniques are labor intensive, time consuming, and difficult to scale. Due to the massive player-bases of online and esports games, it is difficult to conduct qualitative studies at a scale that would allow researchers to gain generalizable insights about a community as a whole.

Time series analysis has looked at gameplay sequences to examine player behavior in a temporal context. However this work focuses more on player classification and win prediction than strategy analysis \cite{saas2016discovering,yang2016real} and considers no context beyond a basic temporal one. Recurrent neural nets (RNNs) also use action sequences to train models for goal recognition but do not involve game context \cite{min2016generalized,min2016player}. Further, these techniques have yet to be used extensively in a domain as complex and dynamic as esports, with attempts to do so often involving simplified versions of the games \cite{bisson2015using}.

Previous work has addressed this issue with the development of \emph{Interactive Behavior Analytics} (IBA), a mixed methods approach to game data analysis that allows for analysts to preserve gameplay context through action sequences and visualization systems \cite{labelPaper,nguyen2015glyph}. The approach was successful in allowing a human-in-the-loop to draw meaningful conclusions about players' strategies, problem solving, and reactions to situations in a dynamic game environment \cite{labelPaper}. However, in IBA the labeling is not informed by the data, rather by experience and what the users see in the spatio-temporal visualization of the action. If the game does not have a spatial component the labeling process will have to be done entirely by hand. Even with the spatio-temporal visualization, users still have to come up with labels based on their experience and the matches they have reviewed. Therefore, inter-rater reliability may not be as good as it can be. In this paper we discuss the use of sequence analysis as a way to allow analysts to observe patterns in the data that can inform and drive the labeling process. In doing so, we constrain the space of possible labels, and streamline the derivation process.

Sequence Analysis (SA) is an established methodology used in the social sciences for discovering characteristics and patterns in data based on sequences of actions or events \cite{ritschard2018sequence}. Action sequences play key roles in popular quantitative techniques such as Recurrent Neural Networks (RNNs) and Time Series Analysis \cite{hooshyar2018data}. There is also a significant body of work exploring ways to enable knowledge discovery and player modeling through SA \cite{chen2015modeling, cavadenti2016did,makarov2017predicting}. The work presented in this paper expand the techniques available by providing a new approach to research data analysis. Specifically, an approach that facilitates context aware, data driven gameplay analysis. This allows analysts to represent and analyze behavioral data in ways that reflect players' strategic thinking and better draw meaning and actionable insight from the data. Being able to do so is especially important for games used in high impact domains, such as education and training.

The rest of this paper will cover a review of existing approaches to game data analytics, IBA, and an overview of sequence analysis. Then, we present an overview of the proposed combination of the two techniques, and a case study in which we demonstrate how this technique allows for the contextualization and analysis of player data from the popular multiplayer online battle arena (MOBA) game: \emph{Defense of the Ancients 2} \cite{dota}.

\section{Related Work}

Numerous techniques exist for analyzing player behavior in these massive, multiplayer, online games, allowing analysts to understand player strategies  \cite{kim2015stage,moura2011visualizing,drachen2012guns,ramirez2010player}. Most existing methodologies can be roughly divided between qualitative techniques, which take ethnographic and observation based approaches \cite{ferrari2013generative,kow2013media,steinkuehler2004learning,Kou2014,bardzell2008blissfully}, and quantitative techniques, which mainly focus on machine learning and statistical analysis  \cite{ong2015player,pobiedina2013successful,stafford2017exploration,nascimento2017profiling,makarov2017predicting}. In this section we will discuss these techniques to provide a background on the current state of the art of game data analysis.

\subsection{Qualitative Approaches to Gameplay Analysis}

Qualitative techniques are beneficial for extracting context, effectively providing the \emph{why} behind a behavioral pattern. While qualitative methodologies (e.g. interviews, observation, and ethnography \cite{steinkuehler2004learning,bardzell2008blissfully}) are effective in providing context regarding the gameplay environment, they are not easily scalable. Due to the massive player bases for online games, it is critical that developers and analysts have access to a methodology that can be applied to large numbers of players quickly and efficiently.

Qualitative work has approached analysis through a variety of different techniques. Ferrari proposed close reading as a methodology for understanding the conventions, tension, strategy, and tactics of a \emph{League of Legends} (LoL) \cite{lol} game, as they are understood by the players \cite{ferrari2013generative}. Ethnographic methodologies are often used in the qualitative study of player behavior: Nardi and Harris used them extensively to learn how \emph{World of Warcraft} (WoW) \cite{wow} players organize themselves within the social and cultural structure of the game to coexist and collaborate effectively \cite{nardi2006strangers}. Steinkuehler performed an ethnographic study within the Massive Multiplayer Online Role Playing Game (MMORPG) \textit{Lineage} \cite{lineage}, in which she identified how the gameplay community educated new players on both the mechanical and social/cultural elements of the game \cite{steinkuehler2004learning}. Structured and semi-structured interviews are among the most common techniques for the qualitative study of online games, as demonstrated by Hussain and Griffiths who conducted 71 interviews with online gamers and identified themes and elements of the online gaming experience \cite{hussain2009attitudes}. Kou and Gui used them to examine elements of social interaction within LoL's temporary teams \cite{Kou2014} and Prax used them to study guild leaders' leadership styles in WoW \cite{prax2010leadership}. Kow and Young used interviews to perform an extensive examination of the key roles \emph{StarCraft's} \cite{starcraft} internet communities play in how new players come to learn the game \cite{kow2013media}. Freeman and Wohn used them to identify patterns and strategies for team formation and communication among esports players \cite{freeman2019understanding}. Combinations of techniques are not uncommon as can be seen in the work of Bardzell et al., who combined ethnographic observation, interviews, and analysis of chat and video logs to examine what players of WoW \cite{wow} considered to be the elements of a ``good" instance run \cite{bardzell2008blissfully}. 

All of these works shed light on the complex human processes that are involved in how players learn interact with each other and work together, especially in online games. However, the qualitative methodologies employed in these works are difficult, if not impossible, to perform at the scale necessary for massive multiplayer online games like esports. Quantitative methodologies present a solution by providing data driven approaches to player behavior and strategy modeling, recognition, and prediction.

\subsection{Quantitative Approaches to Gameplay Analysis}

On the quantitative side of the spectrum there are numerous techniques that focus on identifying and extracting patterns from big data through machine learning, data mining, and statistical analysis \cite{skinner2019artificial,hooshyar2018data}. Many of these methodologies focus on player roles and predicting match outcomes \cite{makarov2017predicting,nascimento2017profiling}. While the existing quantitative techniques are effective at analyzing gameplay at large scales, with the exception of a few works that leverage visualization, much of the context of the game is lost. They effectively identify patterns from data, telling analysts \emph{what} occurred, but often fail to illustrate \emph{why} it occurred. Lacking this context can make it difficult to understand player strategies, especially if one wants to understand them in alignment with how the player-base thinks about them \cite{ong2015player}.

\subsubsection{Player Classification via Quantitative Techniques}

Clustering has often been used to classify player types in online games based \cite{drachen2012guns,ramirez2010player}. Drachen et al. presented an ``in the wild" approach to clustering game data where they integrate knowledge from the design of the game during feature selection and analysis, in order to identify behavioral classes \cite{drachen2012guns}. Ramirez et al. used a meta-clustering approach (using three layers of analysis) to classify player types \cite{ramirez2010player}. Other work has looked at strategies and behaviors in cooperation with, or in place of, player type and role classification, often in the context of win prediction. Eggert et al. used supervised machine learning to identify behavior, such as early movement or ganks, that was indicative of the different player roles \cite{eggert2015classification}. Cavadenti et al. presented a data mining methodology for examining strategic patterns and identifying deviant behaviors in DotA 2 \cite{cavadenti2016did}. Lee and Ramler used a support vector machine to identify player roles in \emph{League of Legends} (LoL) \cite{lol} based on item and spell choices. They used these to identify unconventional team composition strategies and evaluate their success rates compared to conventional teams \cite{lee2017identifying}. 

While clustering is an effective way of quickly identifying and differentiating player types based on in-game behaviors \cite{ong2015player}, the data used tends to focus entirely on aggregated numbers \cite{ong2015player,drachen2012guns,ramirez2010player}. As a result, behaviors are largely removed from the contexts in which they occurred. Even in Drachen et al.'s case, where design knowledge was integrated in the feature selection process, the focus was on aggregated counts such as number of kills, level, looted items, and total playtime \cite{drachen2012guns}. While Ramirez et al. did look at sequence analysis in one step of their multi-step clustering process, they only considered the sequence of coordinates in order to classify players based on exploration habits \cite{ramirez2010player}.

\subsubsection{Data Driven Prediction Techniques}

Win prediction is another goal commonly pursued via quantitative techniques such as machine learning. Yang et al. presented a data-driven method for identifying combat patterns that could lead to victory in MOBA games by converting gameplay data into graphs and analyzing them with established graph metrics algorithms \cite{yang2014identifying}. Schubert et al. derived algorithms for breaking matches down into spatio-temporal encounters in order to examine strategies and identify metrics that could predict match outcome \cite{schubert2016esports}. Both works examine combat encounters and, through mathematical and graph-based metrics, identify factors that can be used to predict victory, such as player positioning, experience gain, gold gain, and kill differences between teams \cite{schubert2016esports,yang2014identifying}. 
Win prediction has also been examined from the perspective of hero selection and role fulfillment. Semenov et al. tested the abilities of Naive Bayes classifiers, Logistic Regression, and Gradient Boosted Decision Trees in predicting DotA 2 match outcomes based on hero picks during the draft phase \cite{semenov2016performance}. Kinkade et al. developed win predictors based on identifying hero roles and how strong a hero was in a given role based on gameplay metrics \cite{kinkade2015dota}.

\subsubsection{Statistical Gameplay Analysis} In addition to machine learning, statistical techniques are commonly used in quantitative approaches. Following the trend of win prediction work, Pobiedina et al. used statistical analysis on DotA 2 log data to examine what aspects of friendship, leadership, and role distribution most significantly contributed to a team's chances of victory \cite{pobiedina2013successful}. Eaton et al. used statistical analysis on LoL data to identify ``critical" team members, team members whose presence or absence has a substantial effect on team performance, and patterns regarding these team members that were strongly related to victory \cite{eaton2017carrying,eaton2018attack}. 

Stafford et al. used statistical analysis to identify patterns of when and with whom \emph{Destiny} \cite{destiny} players played that were indicative of skill acquisition rates and performance \cite{stafford2017exploration}. In all cases, the work identifies elements that can lead to competent and successful gameplay. Eaton et al.'s work, specifically, identified how major gameplay events, events that can heavily influence the outcome of the game, such as teamfights or tower destruction, correspond to the presence and experience of the ``critical" team member, and identify gameplay states based on the status of that member \cite{eaton2017carrying,eaton2018attack}.

\subsubsection{RNN Approaches} Another common methodology for data driven analysis is the use of recurrent neural networks (RNNs), which are also commonly used in traditional sporting environments \cite{bosch2018predicting}. Min et al. have examined the value of Long Short-Term Memory (LTSM) models and identified them among the best approaches to machine driven goal recognition \cite{min2016generalized} as they tend to make more accurate predictions on game data than competing approaches \cite{min2016player,min2017multimodal}. Similarly, Bisson et al. found that their RNN model for plan recognition, which used sequences of observations, performed better than existing models for goal recognition in \emph{StarCraft} \cite{starcraft} \cite{bisson2015using}. Summerville et al. tested RNNs' value in predicting draft picks in DotA 2 \cite{summerville2016draft}. They found that, as the draft went on, RNNs outperformed Bayes Nets in predicting the next hero pick or ban \cite{summerville2016draft}. These techniques involve training models on sequences of actions, but look more for patterns and repetition rather than context. Further, they are more focused on prediction than strategy analysis, and focus on less complex game spaces than esports, with even the work on \emph{StarCraft} \cite{starcraft} being simplified \cite{bisson2015using}.

\subsubsection{Addressing Context}

Some quantitative techniques have attempted to incorporate more context, such as sequential and time series analysis, which preserves temporal context. Saas et al. combined clustering with time series analysis \cite{saas2016discovering} to discover player patterns in free-to-play games. They utilized player data regarding actions taken at a daily frequency, and clustered the temporal sequences based on how different or similar they are from each other \cite{saas2016discovering}. They are then able to analyze these clusters in order to classify player types and predict churn rates \cite{saas2016discovering}. Musabirov et al. combined time-series cross correlation with topic modeling to examine how in game events encourage or discourage conversation topics in live streaming chats \cite{musabirov2018event}. Drachen et al. also combined clustering with time series analysis to examine how in-game behavior relates to the skill level of teams \cite{drachen2014skill}. They focus on clustering players based on spatio-temporal data from DotA 2 and discover patterns in distance between players and movement as they relate to a team's overall skill level \cite{drachen2014skill}. In addition to clustering, Yang et al. used time series data to train a model for win prediction in DotA 2 \cite{yang2016real}. They trained their model on data recorded at time points during gameplay and compare it to prediction based on data collected prior to gameplay. They ultimately found that a combination of the two predicts the outcome with the highest accuracy \cite{yang2016real}. Time series approaches to data analysis allow for the preservation of temporal, and in some cases spatial, contexts. However, the existing work focuses on how this approach can be used to classify players or predict victory, rather than on leveraging the contextual nature of temporal data to analyze player strategies.

Another collection of work has sought to leverage visualization towards the preservation of spatial context \cite{moura2011visualizing,wallner2013visualization,wallner2012spatiotemporal,wallner2019aggregated,labelPaper,van2019modata}. Moura et al. proposed a system that allows analysts to interact with telemetry data, specifically different actions, superimposed atop the game map \cite{moura2011visualizing}. They argued that this technique provides a better understanding of cause and effect than a traditional heat-map \cite{moura2011visualizing}. Wallner et al. used similar approaches in their work, presenting a number of visualization tools that allow for data analysis within the spatial context of the game map both as a temporal sequence \cite{wallner2012spatiotemporal} and in an aggregated manner \cite{wallner2019aggregated}. Evaluations of this approach have found that it is effective in preserving a greater deal of the situational context than other quantitative techniques, therefore increasing the analyst's ability to draw meaning from the data \cite{halabi2019assessing,van2019modata}. However, there is still no consistent, feasible, methodological approach that can deal with context at scale in a standardized manner.


\subsection{Mixed Methods Approaches to Gameplay Analysis}

There is a growing body of work that explores player behavior and strategy via mixed methods techniques. Horn et al. collected data from children playing an educational game and analyzed it using various cluster analysis techniques on sequences of player action \cite{horn2016opening}. The results of this quantitative analysis were combined with the results of a qualitative analysis of think aloud data also collected during the play sessions \cite{horn2016opening}. Canossa et al. also used a mix of quantitative (data mining) and qualitative (direct observation) techniques to create a computational model of player frustration that could be used to detect when players might quit a game \cite{canossa2011arrrgghh}. Mirza-Babaei et al. combined observation based techniques with biometric measures and found that both approaches excelled at exposing different components of the user experience, and that including bio-metric measurement in usability studies can reveal more issues than observation alone \cite{mirza2011understanding}. This work represents a growing interest in mixed methods approaches to player research. However, existing work has largely approached mixed methods research by applying a quantitative and a qualitative techniques separately to different data sets as a part of the same research project. Very little work has explored the qualitative analysis of quantitative data. 

It is clear that there is a need for more mixed methods approaches to game data analytics \cite{zammitto2014player,lisk2012leadership,labelPaper}, especially for ones that can examine gameplay data at scale, while preserving context. In the next section we will provide a concise background on \emph{Interactive Behavior Analytics} (IBA) \cite{labelPaper} and sequence analysis (SA), and then outline our proposed technique to combine the two to create a scalable, generalizable, context sensitive, data driven gameplay analysis methodology. 

\section{Data Driven, Context Aware Gameplay Analysis}

In order to address the drawbacks of IBA cited in previous work \cite{labelPaper}, we propose a methodology that combines \emph{StratMapper}, one of the tools used as a part of the IBA methodology, with sequence analysis. SA benefits the methodology by allowing researchers to see patterns in the data, and use those patterns to inform label derivation. This allows the labels to better represent the data, and increases the chances of achieving a high inter-rater reliability. Although the work presented here uses \emph{StratMapper}, the process can also be applied to a game with no spatial component, by applying the labels derived at the SA stage to the data algorithmically. The results could then be analyzed through another visualization system, such as \emph{Glyph}, the other tool that is used in the IBA technique \cite{labelPaper}.

\subsection{Sequence Analysis}

\emph{Sequence Analysis} (SA)  involves breaking behavioral data into ordered sequences of actions, allowing researchers to see how events unfolded in a temporal manner and in relation to one another. The use of SA spans from psychology, where it has been used to study human perception, cognition, and development, to economics, where it is used to examine operations and markets \cite{abbott1995sequence}. Unlike many of the techniques discussed previously, \emph{Sequence Analysis} involves elements of the context from which the data comes and human analysts who examine the data, rather than mathematical models. Many methods exist for how one approaches the analysis of such data and one may choose to look at sequences in their entirety or search for common sub-sequences within the data \cite{abbott1995sequence,ritschard2018sequence}. As a methodology it is data driven, context aware, and scalable, allowing for its use in many varied situations. Existing work \cite{valls2015exploring,cavadenti2016did,chen2015modeling,bosc2013strategic,canossa2018like} has demonstrated how SA can give designers and researchers a lot more insight than the current aggregate analysis presented and used in the industry and academia. 

\subsection{Interactive Behavior Analytics: IBA}

\emph{Interactive Behavior Analytics} is a mixed methods approach to data driven gameplay analysis that utilizes two visualization systems and a human-in-the-loop to gain meaningful insights regarding player strategies from game data \cite{labelPaper}. The methodology begins with an initial visualization system, \emph{StratMapper}, that displays raw, low level game data in a spatio-temporal context superimposed atop the map of the game from which the data comes, see Figure \ref{fig:FreqBySeg}. This allows analysts to see when and where during gameplay each event was recorded. \emph{StratMapper's} interface includes a timeline that allows users to move forward and backward in time and see how events progressed. It also marks when events occurred in time and what events those were. This essentially allows analysts to ``watch" a replay of the game, without actually watching one. In \emph{StratMapper}, segments of gameplay events can be highlighted on the timeline and labeled by domain experts, and these labels can be exported to the second visualization system \emph{Glyph} \cite{labelPaper,nguyen2015glyph}. \emph{Glyph} provides a visualization of the applied labels as sequences of behaviors, with each behavior represented as a node in a node-link diagram, and the links between each of these nodes representing transitions between states. It also displays a graph of sequence clusters that uses Dynamic Time Warping to visualize how similar or different each sequence is \cite{nguyen2015glyph}. This allows analysts to examine and study behaviors in a temporal context and in relation to each other, but does not include the spatial component of \emph{StratMapper} \cite{nguyen2015glyph}. The procedure of IBA involves using the context provided by \emph{StratMapper} to derive behavioral labels that comprehensively represent what is occurring in the data. This is followed by exporting the labeled data into \emph{Glyph} and then analyzing the sequences to derive conclusions about game context and player strategy. IBA suggests using an iterative process of defining, applying, and testing the \emph{Glyph} visualizations of labels in order to ensure reliability \cite{labelPaper}. 

\subsection{Combining SA and IBA: A new methodology}

While existing work using IBA is beneficial in providing a human centric, data driven approach to gameplay analysis, the labeling process suffers from having too large a possibility space for label generation. Specifically, because existing work using IBA involved deriving labels from only domain knowledge, the number of possible labels was as vast as the number of strategic behaviors that existed within the game. This complicated the process of achieving a suitable inter-rater-reliability. In this work, we attempt to address this drawback of IBA by proposing an alternative approach, that combines sequence analysis with the \emph{StratMapper} portion of IBA in order to streamline the label development by confining the label possibility space to only those behaviors present within the data. 
This combined methodology consists of three steps: 


\begin{itemize}
    \item \textbf{Sequence Analysis Step}: visualizations of player action sequences facilitate the development of preliminary, high level behavioral labels.
    \item \textbf{Refinement, Contextualization, and Reliability Step}: \emph{StratMapper} is used to refine the labels and generate contextual tags through spatio-temporal visualization. Inter-rater reliability (IRR) is conducted to ensure the labels are reliable.  This step may be repeated several times depending on the results of the IRR calculation.
    \item \textbf{Label Application and Analysis Step}: finalized labels are applied across the data set, and those applications are analyzed to draw conclusions about player strategy, behavior, and game context. This can be done in \emph{StratMapper} or other software, such as \emph{Glyph} \cite{labelPaper,nguyen2015glyph}.
\end{itemize}

\subsubsection{Sequence Analysis Step}

The initial procedure in the combined methodology is to transform the data such that it can be represented as sequences of abstracted actions for each player. Although IBA will eventually have domain experts apply labels to the data, at this stage it is still important to ensure that the sequences are human interpretable. Thus, the raw data must be put through an abstraction process in which higher level behaviors are derived from primitive actions. This can be done in a data driven manner at this stage. Labels can be derived from information drawn directly from the data, such as proximity between players. Thus, domain experts are not required to apply labels manually at this stage. However, it is important that the analysts keep in mind the types of strategic behaviors they plan on examining when they decide on this abstraction, as it is important that the abstracted data provide the right context to derive initial labels. Displaying the abstracted sequences can be done using the analysts visualization of choice.

Once the sequences are abstracted and visualized, analysts who are familiar with the domain can analyze the sequences. The goal of this step is to look for frequent patterns within the sequences and sub-sequences and derive behavioral meaning from them. Through this process, a list of high level labels is derived from the sequence data. These labels capture possible behaviors or strategies that were performed by the players represented by the data. Observations regarding the temporal occurrences and frequencies of certain patterns can be recorded at this stage.

\subsubsection{Refinement, Contextualization, and Reliability Step}

Once the high level labels are derived, the analysts can use \emph{StratMapper} to refine and contextualize the labels. This is done through an iterative process in which the temporal occurrences of sequence patterns in the SA step are used to direct analysts to certain time points in gameplay within \emph{StratMapper}. Here, analysts can use the \emph{StratMapper} toolkit (timeline, filtering of events, etc.) to examine what players were doing, and see if their observations of player behavior match the labels created during SA. Based on these observations, labels can be adjusted and combined to better represent player behavior. Further, \emph{StratMapper} facilitates the observation of multiple players and non-player characters interacting in a spatio-temporal context, allowing the analysts to examine what was happening before, during, and after a player engaged a given behavior, and derive contextual information. 

The derived contexts can be refined into contextual tags that can be combined with the refined behavioral labels to create a rubric of ``behavior" - ``context" combinations that are applicable to the game data at large. For reliability, it is recommended that the rubric be refined through an iterative process: two, or more, researchers can apply the ``behavior" - ``context" combinations separately, and then an inter-rater reliability measure can be calculated to determine if there is agreement regarding the meanings of each label and tag. If the agreement is not sufficiently high, the researchers can re-convene to examine and refine the rubric, and repeat the process until agreement is achieved.

\subsubsection{Label Application and Analysis Step}

When the rubric achieves sufficient reliability, analysts can use \emph{StratMapper} to apply the behavioral labels and contextual tags across as much gameplay data as they desire. Upon completion of the labeling process, the applications of the ``behavior" - ``context" combinations can be analyzed to draw conclusions about player behavior. This can be done either in \emph{StratMapper} itself or by exporting the labels to another analysis tool, such as \emph{Glyph} \cite{labelPaper,nguyen2015glyph}. The nature of the analysis depends on the game, and the questions that analysts hope to answer. Some example analyses are the frequency of given ``behavior" - ``context" applications at different points in gameplay, the frequency of certain labels or tags to applied to certain character types, and the frequency of certain combinations to occur in succession.

\section{Case Study: DotA 2}

To demonstrate the utility and benefits of the proposed methodology, we conducted a case study in which we used the technique to perform context sensitive analysis of DotA 2 players grouping and soloing behaviors using log data to gain insights into how players interact (or do not) and how that changes over the course of a game. 

\subsection{The Game}

A game of DotA 2 pits two teams, the ``Radiant" and the ``Dire", of five players against each other in a capture-the-flag style competition. Each team has a base, located on opposite ends of the game map, containing an entity known as the ``Ancient". The goal of the game is to protect one's own Ancient while simultaneously trying to destroy the opposing team's. The game map is organized into three lanes with forested areas between them, populated by various NPC (Non Player Character) entities ranging from towers that fire at enemy players, to monsters that can be slain for experience and gold. An average DotA 2 match can last between 30 minutes to upwards of an hour, and is a highly dynamic environment in which players will constantly transition between solo and team play as they pursue objectives. Matches are usually split into three phases (early, mid, and late game) separated by significant changes to the state of gameplay. Though the exact transition periods are not strictly defined, early game typically ends when one team has successfully destroyed one of the enemy's towers, and mid game typically transitions into late game when one team has successfully destroyed enough towers to reach the entrance to the enemy base. 

Each player in a DotA 2 match controls an avatar called a ``hero" and their responsibilities are determined by which hero they chose to play. DotA 2 heroes are divided into categories based on their skills and how they contribute to gameplay. Some common hero types include ``carries": damage oriented heroes with low defense who start vulnerable but scale with each level into powerful killing machines; ``supports": defense and healing oriented heroes who do not deliver much damage but possess skills that can protect their teammates; and ``initiators": heroes who possess skills that allow them to quickly cross distances and incapacitate targets, making them excellent at setting up fights and kills. Success in a DotA 2 match is often determined by a team's ability to select heroes that synergize well together, and fulfill their roles correctly \cite{schubert2016esports,pobiedina2013successful}.

\subsection{Process}

\subsubsection{Data Abstraction and Representation}
For this case study, we are using data logged from approximately 200 DotA 2 players. For the analysis, each player's sequence was split into multiple sequences based on game segment. As stated above there is some ambiguity in regards to when a game segment begins and ends. For our purposes, we mark the end of early game as the moment when the first tower falls, and the end of mid game when the first tier-3 tower (the tower closest to a team's base) falls. Not all games make it to late game, as teams are allowed to surrender, thus some players only had two segmented sequences. After removing the sequences that came from games that never reached late game, there were 450 sequences divided evenly across the three game segments (150 sequences in each segment). For our purposes, each player's primitive actions are abstracted based on proximity to allied and enemy players, as previous work emphasizes the significance of player positioning \cite{schubert2016esports,drachen2014skill,yang2014identifying} and we are interested in exploring how players behave in teams in terms of grouping or splitting up. A player was considered to be “nearby” if they were within 81.92 units of the observed player. The list of abstracted behaviors was: 

\begin{itemize}
    \item ``Solo": a player has no allies in their vicinity, indicating independent gameplay such as farming or roaming.
    \item ``Fight": a player has encountered at least one enemy player, indicating that they are likely engaging in combat.
    \item ``Kill hero": a player kills an opponent.
    \item ``Teaming": a player has allied players in their vicinity, indicating that at least a few members of a team have chosen to work together rather than independently.
    \item ``Death": a player dies.
    \item ``Harassed by opponents": a player encounters multiple enemy players perhaps because the player is a threat being focused or they have fallen into a trap.
    \item ``Fight diminishes": the number of opponents in a player's vicinity decreases, indicating that a would-be fight is breaking up or that players are fleeing.
    \item ``Fight intensifies": the number of opponents in a player's vicinity increases, indicating that a minor encounter is growing in intensity and may escalate to a team fight.
    \item ``Team fight": a player has more than one ally and opponent in their vicinity, indicating the possibility that a team fight is occurring or about to occur.
    \item ``Full team assembly": a player's entire team is in their vicinity, the most likely explanations for this are team fights and pushing objectives.
\end{itemize}

\begin{figure*}[ht]
    \centering
    \subfigure[]{
    \includegraphics[width=.3\textwidth]{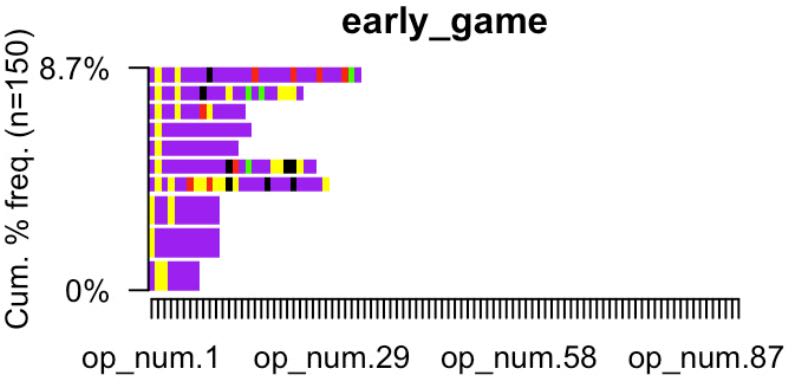}
    \label{fig:earlySeq}
    }
    \hlinesep
    \subfigure[]{
    \includegraphics[width=.3\textwidth]{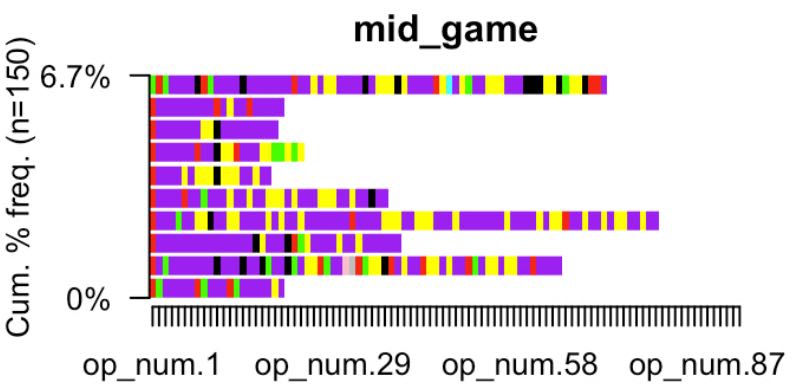}
    \label{fig:midSeq}
    }
    \hlinesep
    \subfigure[]{
    \includegraphics[width=.3\textwidth]{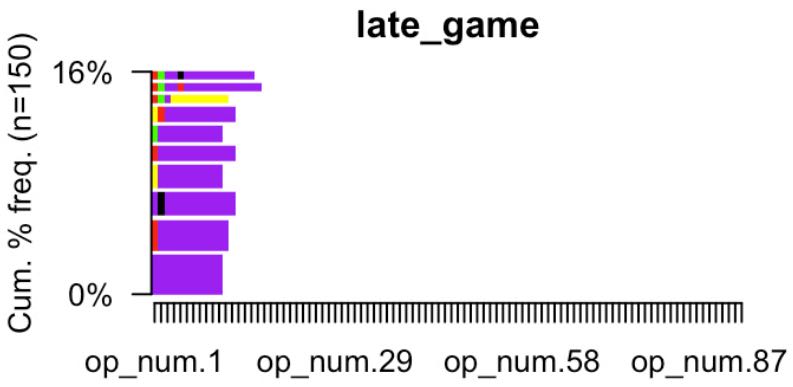}
    \label{fig:lateSeq}
    }
    \subfigure[]{
    \includegraphics[width=.9\textwidth]{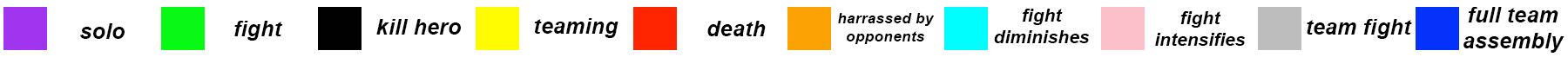}
    \label{fig:key}
    }
    \caption{The most frequent sequences in each game segment, frequency on the y axis and operation number on the x axis, and the key in (d).}
    \label{fig:SeqBySeg}
\end{figure*}

For our purposes, we use the \emph{TraMineR} R package \cite{gabadinho2011analyzing} to visualize the abstracted sequences discussed above. In Figure \ref{fig:SeqBySeg} we present the TraMineR output for the ten most frequent sub-sequences in each game segment (early, mid, and late). The y axis represents the frequency, with the numbers indicating what percentage of the overall sequences are represented by these most frequent, and the height of a given sequence representing the percentage of that amount represented by that sequence. The x access represents operation number, in other words, how many actions were in a sequence, and where in that a given action falls. Long bands of a single color represent one behavior occurring in succession.
 
\subsubsection{Deriving Labels and Tags}

By analyzing the \emph{TraMineR} sequences, three researchers developed an initial list consisting of ten high level behavioral labels, and developed a general idea of when during gameplay certain labels were more likely to occur. The ten labels were as follows:

\begin{itemize}
    \item ``Team Kill": a team works together to kill an opposing player, derived from the action pattern where teaming states occurred in frequent succession with kill states. 
    \item ``Team Recovery": a player sticks close to team mates after being killed, derived from the action pattern where teaming states appeared in frequent succession after an occurrence of one or more death states.
    \item ``Teaming": a high level behavioral label for instances of teammates staying near each other, derived from the action pattern of teaming states in succession. 
    \item ``Soloing": a high level behavioral label for instances of a player being alone, derived from solo states in succession.
    \item ``Solo Killing": a player is able to secure kills on their own without the aid or protection of teammates, derived from frequent soloing states interrupted by kill states.
    \item ``Solo Recovery": a player stays on their own after dying, derived from solo states in frequent succession after one or more occurrences of a death state.
    \item ``Focus Target": a player is focused by the enemy and frequently attacked or killed, derived from frequent kill states followed by frequent death states.
    \item ``Team Fighting": the team fights together as a group, derived from teaming states in succession with kill and fight.
    \item ``Assist": a player aids another player in achieving goals but does not necessarily achieve the goals them self, derived from frequent teaming states with death and fight states among them, but no kill states.
    \item ``Kill Exchange": two players kill each other or each team loses one player in a team fight, derived from two sequences with similar action patterns leading to a death and a kill in approximately the same location in the sequence.
\end{itemize}

\begin{table*}[]
    \centering
    \begin{center}
    \begin{tabular}{ |c|c|c| } 
     \hline
     \textbf{Label} & \textbf{Tag} & \textbf{Description} \\ 
     \hline
     \hline
     Team Fighting & Objective Struggle & The team fights over an objective such as a tower \\ 
      & Retaliation & The team fights back to get revenge for something the other team did \\ 
      & Focus Target & The team fights in order to take down a particular player \\
     \hline
     Assist & Scout & The player roams the map, laying or removing wards \\
     & Vanguard & The player is out in front, soaking damage or using stuns to keep the enemy at bay \\
     & Rearguard & The player brings up the rear during escape, protecting escaping teammates \\
     & Babysitter & The player ``babysits" players, providing heals and shields\\
     \hline
     Solo Recovery & Farming & The player kills non player entities alone after reviving\\
     & Scout & The player roams the map alone after reviving \\
     & Push & The player moves the center of action closer to the enemy side of a lane after reviving \\
     \hline
     Team Recovery & Push & The player joins others to move the center of action to the enemy side of a lane after reviving \\
     & Objective Struggle & The player joins others to go after an objective, such as a tower, after reviving \\
     \hline
    \end{tabular}
    \end{center}
    \caption{The rubric of behavioral labels and contextual tags before the IRR process}
    \label{tab:iteration1}
\end{table*}

Once the ten behavioral labels were identified, the same researchers used them as a guide while exploring the replay data of a single game in \emph{StratMapper}. Guided by the SA step, the researchers looked for occurrences of the behaviors in the replay data, and then used the tools afforded by \emph{StratMapper} to explore the context surrounding the occurrence. There were two goals to this process, the first was to refine and consolidate the list of labels.

During this procedure, the list of ten labels was refined into a list of four labels. One of the initial labels, ``Focus Target", was converted to a contextual tag for the ``Team Fighting" label, as it was observed that the two behaviors overlapped substantially. The high level labels of ``Teaming" and ``Soloing" were identified to also be too vague and overlap too greatly with the rest of the labels and were thus removed from the set. ``Team Kill" was removed due to overlap with other labels (a team kill always occurred within a team fight). ``Kill exchange" was removed due to difficulty identifying the event within \emph{StratMapper}, and ``Solo Killing" was removed due to it being a behavior that encompassed a longer stretch of time than the others (it implied a player acquiring a number of kills while playing alone for a length of time) and therefore was identified as not matching, conceptually, with the rest of the behavioral labels.

The second goal was to develop contextual tags that could be applied to the behavioral label that could capture the goals and motivations of the players. For example, in the case of the ``Team Fighting" label, the researchers looked for an occurrence of a team fight, and then examined what occurred before, during, and after the bulk of the fight, in order to develop a hypothesis for \emph{why} the team fight occurred. If the team fight was immediately followed by a tower being taken, then the context would be identified as an ``objective struggle" as this was interpreted as the goal of the fight. Thus, ``objective struggle" was derived as a contextual tag that captured player goals. The rubric as it appeared after this first iteration can be seen in Table \ref{tab:iteration1}.

\subsubsection{Inter-Rater Reliability}

In order to test the reliability of these labels and their tags, two of the three researchers performed a reliability measure in which the labels were applied separately to the mid game portion of a DotA 2 match, and Cohen's Kappa was calculated. The resulting IRR indicated promise for the labels and tags surrounding team fighting (IRR = .63), however there was no agreement regarding the other contextual labels (IRR < .1). Thus, the researchers performed a label refinement procedure in which two researchers applied labels together to clear out conceptual confusion. During this process, the ``Assist" label and all of its tags were removed due to agreement that this was a much more low level behavior than the other behaviors being labeled throughout this process. However, a tag for ``Team Recovery" called ``Assist" was defined at this stage, in order to capture players who joined other players after re-spawn. This final list of labels was put through a second IRR measurement resulting in a score of .60 indicating moderate agreement across all labels. The rubric as it existed after this iteration (the final version) can be seen in Table \ref{tab:iteration2}.

\begin{table*}[]
    \centering
    \begin{center}
    \begin{tabular}{ |c|c|c| } 
     \hline
     \textbf{Label} & \textbf{Tag} & \textbf{Description} \\ 
     \hline
     \hline
     Team Fighting & Objective Struggle & The team fights over an objective such as a tower \\ 
      & Retaliation & The team fights back to get revenge for something the other team did \\ 
      & Focus Target & The team fights in order to take down a particular player \\
     \hline
     Solo Recovery & Farming & The player kills non player entities alone after reviving \\
     & Scout & The player roams the map alone after reviving \\
     & Push & The player moves the center of action closer to the enemy side of a lane after reviving \\
     \hline
     Team Recovery & Push & The player joins others to move the center of action to the enemy side of a lane after reviving \\
     & Objective Struggle & The player joins others to go after an objective, such as a tower, after reviving \\
     & Assist & The player helps another player after reviving \\
     \hline
    \end{tabular}
    \end{center}
    \caption{The rubric of behavioral labels and contextual tags after the IRR process}
    \label{tab:iteration2}
\end{table*}

\subsection{Results}

By applying the labels from the final set to a single DotA 2 match in \emph{StratMapper}, we observed patterns and trends in how player behavior and the context behind that behavior varies depending on the gameplay segment and the players involved. In the following sections we will detail some of the patterns that were observed by analyzing the label applications of the lead labeler.

\subsubsection{Team Fight Differences in Early and Late Game}

``Team Fighting" labels were applied 13 times during the early game segments. Of those applications, ten used the ``Focus Target" contextual tag, while the remaining three used the ``Retaliation" tag. In almost all cases of a focus target, the attacks involved two to three attackers against one target, and the target was an offensive carry hero. Many of these attacks occurred within the lane that the offensive hero was positioned in for early game play, and often after that hero had been left alone for a lengthy period of time. This indicates that the primary condition to gather for a team fight during early game was to focus a target, and the primary condition to be focused was to be an offensively oriented hero. By contrast, during late game, after the final tower fell, ``Team Fighting" was applied 17 times, however, the ``Focus Target" tag was only used five times.


A possible explanation for these strategic differences can be found in the design of the game. Carries start off weak but eventually become powerful glass canons. This is especially true if they have been allowed to farm (kill non player entities for gold and experience) un-interrupted for too long. Thus, it is strategically advantageous for players to target offensive heroes early game. If they are able to kill the hero, they can stop them from killing NCPs to obtain gold and experience, effectively cutting them off from their power supply. This would also explain why many of the fights appeared to be occurring within the target's lane. Further, because the carries had yet to become powerful killers, two to three attackers was sufficient. 


However, the results indicate a clear strategic change when the game shifts to late game. Instead of initiating team fights to take down carries, players are now fighting over objectives. Again, an examination of the design of the game can provide a possible explanation. In the late game environment, carries have typically achieved their maximum power, and are unlikely to be farming or performing any task strictly focused on building power. This means it is less beneficial to try and kill them than it is early game. Instead, taking down, or defending, objectives such as buildings or Roshan becomes much more critical, as successful completion of these can turn the tide of battle. Thus, teams are more likely to initiate a team fight with this particular goal during the late game. That being said, there are still cases where a single hero will attempt to push their way towards an enemy base and needs to be dealt with, which is why ``Focus Target" did still appear occassionally.


\subsubsection{Early Game vs. Late Game Goals}

During early game, a targeted carry died during an encounter labeled as ``Team Fighting - Focus Target" four times. ``Solo Recovery - Farming" was applied to those heroes immediately afterwards three times. From these results, we can conclude that it is strategically advantageous for a carry to farm after recovering from an early game death, likely because many of these players were farming when they were killed. 

It is interesting that there is such a preference towards a solo behavior in a team based game, especially when one returns from death weaker than the rest of the team and could benefit from the protection of numbers. However, we can again connect this to elements of the game's design. In DotA 2, gold and experience are shared between all nearby heroes, meaning that effective farming is actually best done in isolation.


``Solo Recovery" labels were applied eight times during early game, while ``Team Recovery" labels were applied two times. However, during late game, ``Team Recovery" appeared nine times, while ``Solo Recovery" did not appear at all. The ``Push" and ``Objective" tags were also used more often late game, seven times, than early game, two times. These changes indicate another strategic shift that occurs when the game transitions from one phase to another. 

Again, knowledge of the design of the game can aid in understanding how the context has influenced players' actions. During late game, when players manage to kill a significant number of the opposing team, they take advantage of the lack of opposition to push objectives. This results in re-spawned players having to focus on regaining lost ground so as not to lose the match, a strategy achieved best by relying on strength in numbers. Thus, ``Farming" becomes a non-ideal recovery goal and instead ``Team Recovery" becomes more prominent as players focus on taking objectives.


\subsubsection{Frequency of Behavior Across Gameplay}

\begin{figure*}[ht]
    \centering
    \subfigure[]{
    \includegraphics[width=.3\textwidth]{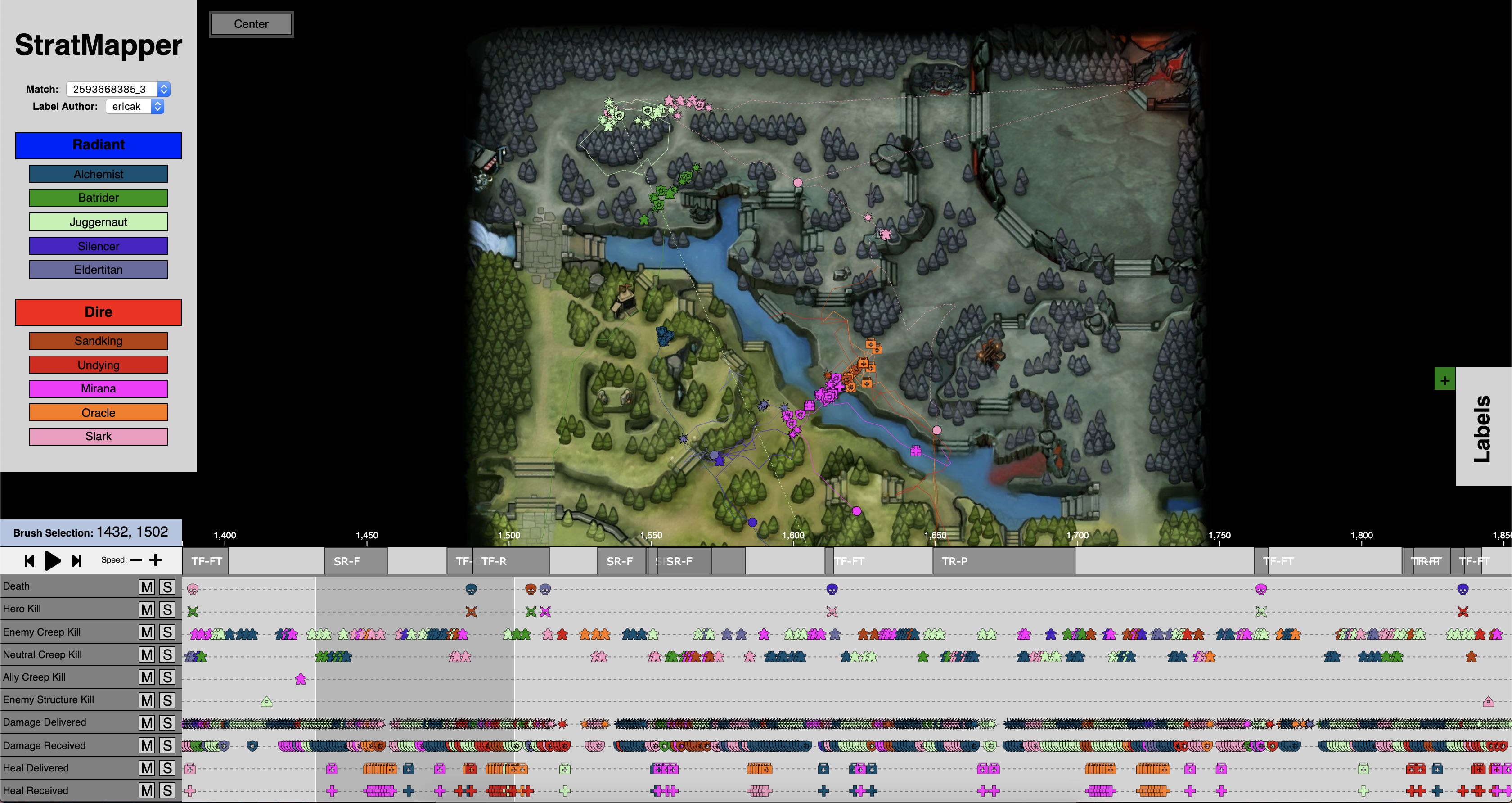}
    \label{fig:early}
    }
    \hlinesep
    \subfigure[]{
    \includegraphics[width=.3\textwidth]{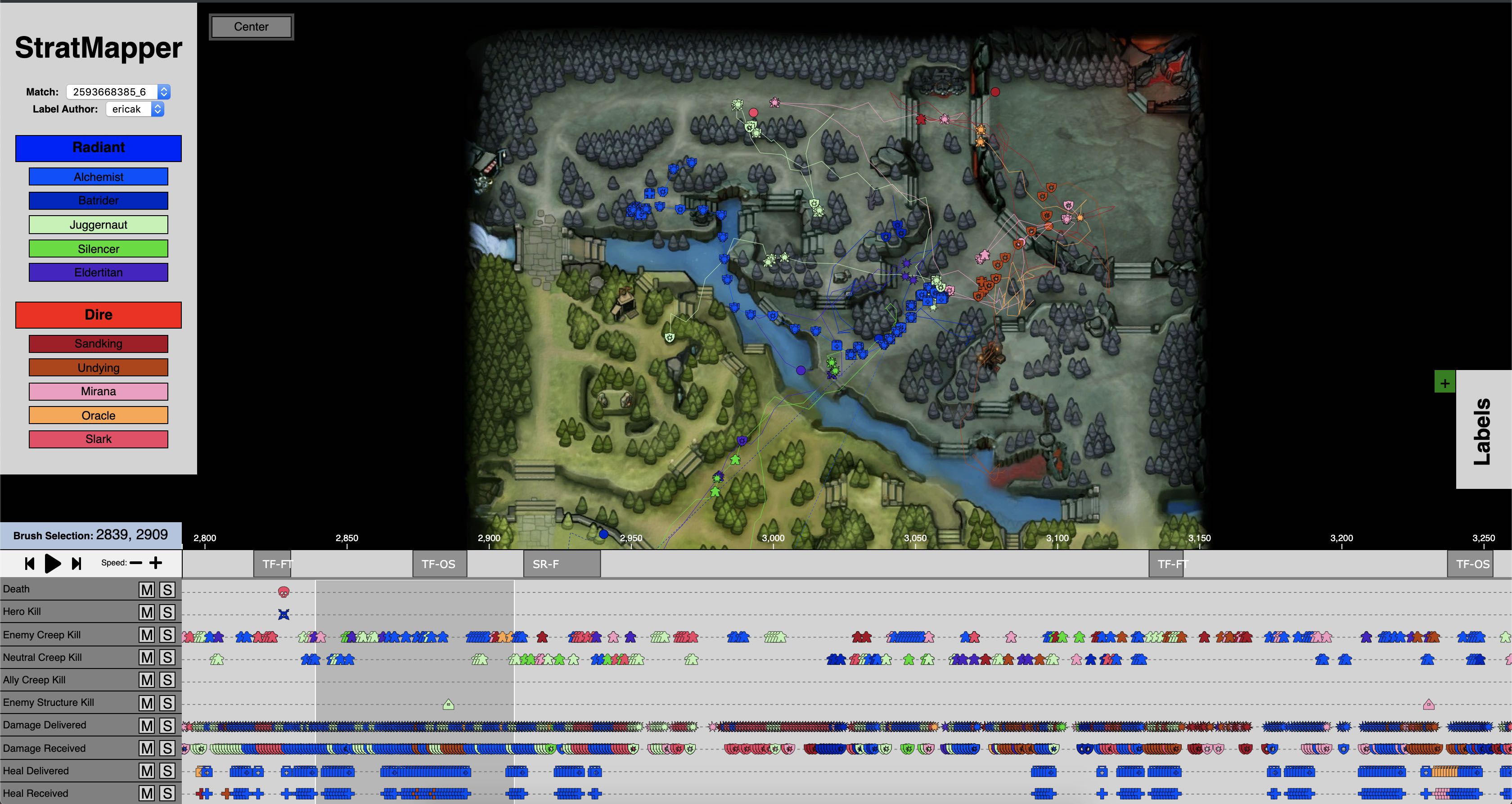}
    \label{fig:mid}
    }
    \hlinesep
    \subfigure[]{
    \includegraphics[width=.3\textwidth]{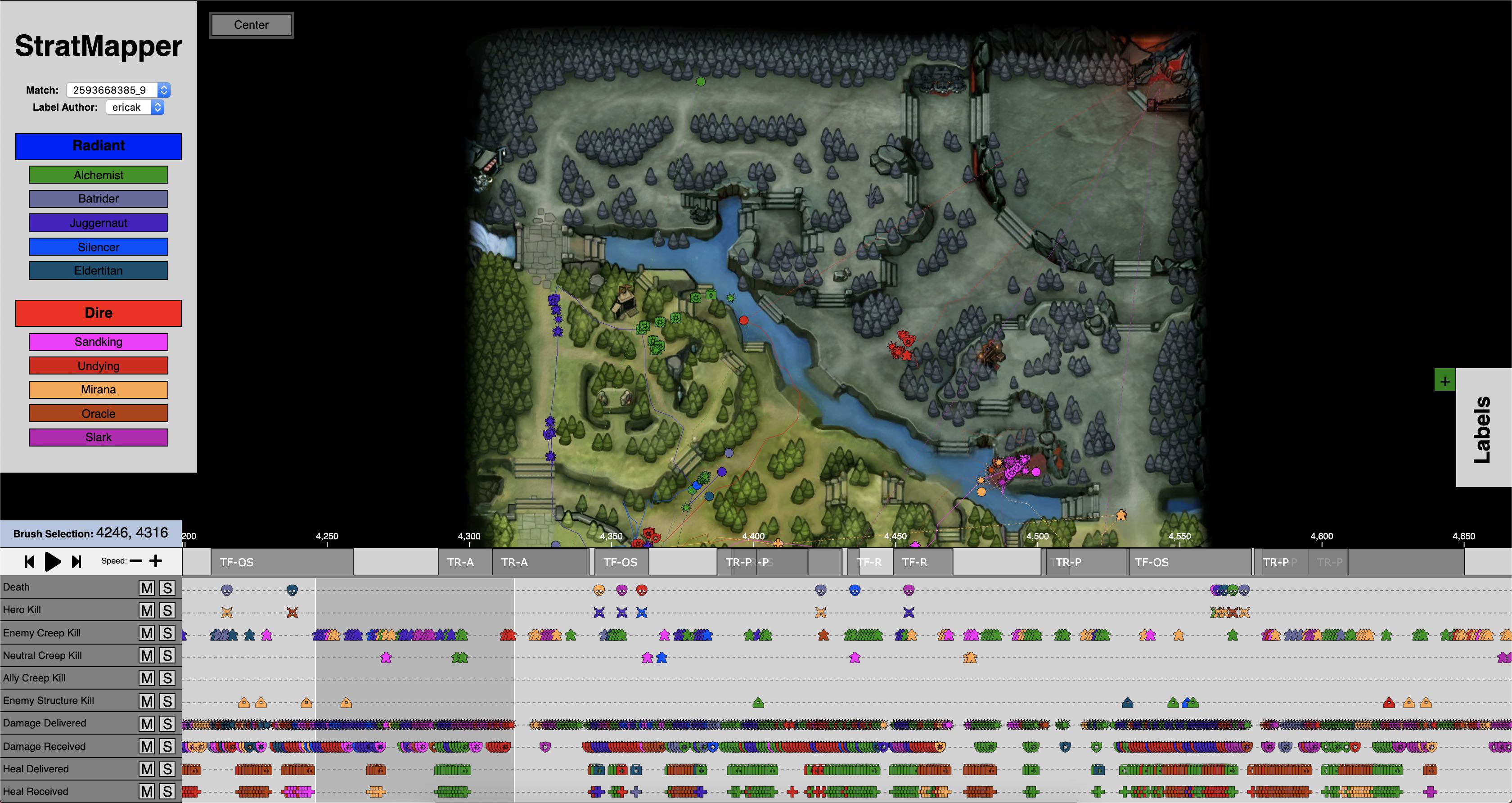}
    \label{fig:late}
    }
    \caption{Distribution and Frequency of labels during early (a), late mid (b), and late game (c). The timeline in (b) consists of notably fewer labels than (a) and (c), indicating a drop in activity related to player grouping.}
    \label{fig:FreqBySeg}
\end{figure*}

``Team Fighting" labels, with various tags, were applied fairly regularly during early game, 13 applications, and late game, 17 applications. However, it was relatively sparse during the latter parts of mid game, four applications. This pattern can be seen in Figure \ref{fig:FreqBySeg}. The reasoning behind this strategic pattern can be explained, once again, by examining the game design of DotA 2 and the ways that players and their behavior are influenced by it.

As a game of DotA 2 progresses, the amount of time between a death and a re-spawn increases dramatically. During latter parts of the game, this downtime is significant enough to turn the tides of a match. Once a team's tier-3 tower falls (the moment we define as the beginning of late game), the enemy has relatively easy access to their base, putting them in a perilous position, and towers are most easily taken when members of the defending team are dead. It was observed during both the SA and \emph{StratMapper} steps that game transitions typically coincided with player deaths. Thus, it can be concluded that the scarcity of team fights during the last moments of the mid game are likely the result of players trying to avoid dying, lest they lose their tower. This sheds light on how players' willingness to engage a strategy at all is influenced by the dynamic context of the game environment.


What is curious though is that the frequency picks back up in late game, despite the fact that the time spent dead does not decrease. There are items that allow players to revive instantly, and it may be that acquiring these means players do not need to be as cautious. It may also be that once the tier-3 tower is lost, players feel that becoming more offensive is necessary. On the losing side, this may be a desperate effort to push the action back to the enemy's side of the map, while on the winning side this may be a focus on trying to finish the match before the losing team can tip the balance. 


\section{Discussion}

Combining IBA and SA results in a new methodology that facilitates context aware knowledge discovery without having to recruit and observe participants. Through the DotA 2 case study, we demonstrated how capturing the context behind players' actions allowed us to link their strategic decisions back to elements of the game's design. Being able to obtain these kinds of insights is applicable to various domains. When dealing with data from games in high impact domains, such as education, it is especially important that researchers and analysts can understand the context behind player actions, such that they can better apply their findings. In the entertainment domain, developers who wish to keep their games relevant and appealing benefit from capturing the context behind player action as it allows them to make more informed design decisions.



Using SA as an initial step for high level label development as a part of the IBA procedure helped streamline the label development process by confining and focusing the possibility space for derived labels. Although previous work reports a higher IRR \cite{labelPaper}, the process of reaching an acceptable number was far more labor intensive. Without SA, labels were derived purely from domain knowledge, resulting in a large list that represented a variety of behaviors that could exist within the data based on the possibility space for actions within DotA 2. In order to feasibly apply the labels, the list had to be reduced in size dramatically, and labels needed to be adjusted and redefined in order to reach an acceptable agreement \cite{labelPaper}. Here, we used SA as an initial step to allow the data to inform the label derivation process alongside domain knowledge. In doing so, we prevented the list of labels from becoming unwieldy, and were able to reach an acceptable agreement in fewer iterations, and with less dramatic changes to the list of labels. We argue that further refinement of the approach in future work would result in higher IRR. Sequence analysis allows the research team to formalize and contextualize the labels. It can quantify behavior to allow for more concrete label generation, a process that can be difficult when working with a game as complex as DotA 2. This results in labels with higher certainty that they correctly represent the behaviors present in the data. While another approach could be to automate the labeling process, this technique is beneficial in that it emphasizes and preserves human judgement throughout the entire methodology.

This approach is generalizable across various game types and genres. First, both SA and IBA have been used in various contexts in previous work \cite{labelPaper,ritschard2018sequence,abbott1995sequence}. Second, all player behaviors in a game can be represented as a sequence of actions. For example, while we use DotA 2 in the paper, sequences of actions in a role playing game (RPG) may take a similar form with different content, e.g., [talk to NPC][move to quest location][pick up key item][etc...]. A developer may wish to know why players of this RPG never complete a given quest. From aggregated data, they may see where in a quest progression the players are giving up, but may not be able to see why. Through the proposed methodology, it would be possible to capture the context surrounding the players' behavior, such as what elements of an environmental layout are making it difficult for players to progress to a necessary location. Such knowledge, of how player behavior is impacted and influenced by the contextual nature of the environment, is invaluable to all genres. Thus, we argue this methodology can be generalized as long as researchers are able to develop appropriate abstractions.

\subsection{Limitations and Future Work}

While we argue that the combined methodology is more efficient than IBA alone, we acknowledge that more research is needed. Future work will examine the extent to which the combined methodology improves upon IBA, and explore ways to further streamline the label derivation procedure.

We also acknowledge that we have only demonstrated its use in a single case study on a single game. To truly ensure generalizability, more work is necessary in which different games are analyzed and various abstraction techniques at the SA step are explored. The road to investigating this will take on many different forms and thus will be the subject of future work. 

Additionally, we argue that the insights gained from such a context aware, human-in-the-loop technique are more beneficial and actionable to developers than those gained through quantitative methodologies, and that it is a more streamlined approach than the previous IBA technique. However, further study and evaluation is required to determine to what extent this is the case and in what ways one approach may be more beneficial than the other. Proper evaluation studies will be a part of future work as will be continued refinement of the combined technique.

\section{Conclusion}

Game data analytics are critical to the future of gaming, especially massive, multiplayer online games and games in high impact domains. If developers wish to keep their player-bases engaged, they must constantly update their games to correct flaws, improve design, and create new and interesting experiences. Similarly, if the creators of educational or training games want to aid those that are struggling, they need to be able to determine what is giving the students trouble. However, traditional methodologies often separate data from context, making it difficult to examine or understand player strategies and problem solving techniques as the players understand them. This can result in uninformed design adjustments that may harm the player experience and survival of the game.

While qualitative methodologies provide context focused approaches, they are difficult, if not impossible, to perform at the scale needed for many popular games. Quantitative methodologies grow more advanced every year, but those that attempt to understand players as they understand themselves are still in their infancy, and are difficult to apply to complex, online games. Previous work \cite{labelPaper} presented the IBA methodology as a way of combining the scalability of quantitative approaches with the context aware analysis of qualitative methods. 


We build upon IBA and develop a combined methodology that uses sequence analysis as a first step to guide the labeling process. Through the analysis of action sequences, analysts are able to develop high level labels, which can then be contextualized and tuned using \emph{StratMapper} to create a rubric of ``Behavioral Label" - ``Contextual Tag" combinations. By developing the labels through SA, the methodology ensures that the labels will represent patterns that are present in the data, restricting the possibility space to only what is relevant. Through a case study with DotA 2, we demonstrate how this approach facilitates the preservation of context, which subsequently allows analysts to draw insights about how the design of a game may be influencing player strategy.

Sequence analysis benefits from keeping the human at the center of the methodology, ensuring that the results are human interpretable throughout the entirety of the IBA methodology. In future work we intend to explore the generalizability and usability of the combined methodology and evaluate it against existing, state of the art techniques.


\bibliographystyle{ACM-Reference-Format}
\bibliography{sample-base}

\end{document}